\documentstyle[preprint,eqsecnum,aps]{revtex}
\tightenlines

\newcommand{\be}{\begin{equation}}
\newcommand{\ee}{\end{equation}}
\newcommand{\bea}{\begin{eqnarray}}
\newcommand{\eea}{\end{eqnarray}}
\newcommand{\ba}{\begin{array}}
\newcommand{\ea}{\end{array}}

\newcommand{\Th}{\Theta}

\newcommand{\de}{\delta}
\newcommand{\pa}{\partial}
\newcommand{\pari}{\partial^{-1}}
\newcommand{\no}{\nonumber}
\begin{document}
\draft
\title{The nonstandard constrained KP hierarchy and the \\
generalized Miura transformations}

\author{Ming-Hsien Tu}

\address{ Department of Physics, National Tsing Hua University,\\
Hsinchu, Taiwan, Republic of China.}

\date{\today}

\maketitle

\begin{abstract}
We consider the nonstandard constrained KP (ncKP) hierarchy which is obtained
from the multi-constraint KP hierarchy by gauge transformation. The second 
Hamiltonian structure of the ncKP hierarchy can be simplified by factorizing 
the Lax operator into multiplication form, thus the generalized Miura transformation
is obtained. We also discuss the  free field realization of the associated W-algebra.
 
\end{abstract}


\newpage

\section{Introduction}
In the past few years, there are several intensive studies on the relationships
between conformal field theory and integrable system  which include, in
particular, exploration of
the role played by the classical $W$-algebras  in integrable systems \cite{BS}.
It's Adler map (see, for example, \cite{D1}) from which the $W$-algebras can be 
constructed as Poisson bracket algebras. 
A typical example is the $W_n$ algebra constructed from 
the second Gelfand-Dickey (GD) structure of the $n$-th Korteweg-de Vries (KdV) 
hierarchy \cite{DIZ,D}. 
Amazingly, by factorization of the KdV-Lax operator, the second Hamiltonian
structure is transformed into a much simpler one in an appropriate space of the
modified variables. 
Thus the factorization  not only provides a Miura transformation which maps the 
$n$-th KdV hierarchy to the corresponding modified hierarchies, 
but also gives a free field realization of the $W_n$ algebra. 
This is what we called  the Kupershmidt-Wilson (KW)
theorem \cite{KW,D2}. The generalization of the KW theorem
to the Kadomtsev-Petviashvili (KP) hierarchy and its reductions have been discussed
\cite{C,BX,BLX,D3,Y,ANP,MR}.
In general, the above scheme is encoded in the particular form of the
Lax operator and its associated Poisson structure. Therefore, 
the number of integrable hierarchies where the KW theorem works is
quite limited. 

Recently, Q P Liu \cite{Liu} conjectured that the above scheme also works
for the constrained modified KP (cmKP) hierarchy\cite{OS}, which is a kind of reduction
of the KP hierarchy. The proof has been given in two recent papers \cite{ST2,L2}
based on the observation \cite{HSY} that the second Hamiltonian structure 
of the cmKP hierarchy can be mapped into the sum of the second and the 
third GD brackets. Therefore one can factorize the Lax operator of the cmKP 
hierarchy into linear terms. In this paper, we generalize the previous 
results \cite{ST2,L2} to the nonstandard constrained KP 
(ncKP) hierarchy \cite{OS}, which is obtained from the gauge transformation of 
the multi-constraint KP hierarchy\cite{OS}. We find that the second Poisson 
structure of the ncKP hierarchy can be simplified by factorizing the 
nonstandard Lax operator into multiplication form containing inverse linear terms.

This paper is organized as follows: In Sec.II we consider the
multi-constraint KP hierarchy. Using the 2-constraint KP hierarchy as an
 example, we calculate its Poisson brackets from its second Hamiltonian structure and
discuss its associated conformal algebra.
Then in Sec. III we perform a gauge transformation
to obtained the nonstandard cKP hierarchy and the corresponding Poisson
brackets. We find that after mapping the nonstandard Lax operator to a 
1-constraint KP Lax operator, the Poisson structure becomes the sum of the
second and the third GD brackets defined by the 1-constraint KP Lax operator.
We also show that the conformal algebra associated with the nonstandard
Lax operator is encoded in the conformal algebra of the 1-constraint KP Lax operator.
In Sec. IV we simplify this Poisson structure by factorizing the Lax operator
into multiplication form and thus obtain the generalized Miura transformation.
Conclusions and discussions are presented in Sec. V.

\section{multi-constraint KP hierarchy}
The multi-constraint KP hierarchy is the ordinary KP hierarchy restricted to 
pseudo-differential operator of the form
\be
L_{(N,M)}=\pa^N+u_2\pa^{N-2}+\cdots+u_N+\sum_{i=1}^M\phi_i\pa^{-1}\psi_i.
\label{multi}
\ee
The evolution of the system is given by
\bea
\pa_k L_{(N,M)}&=& [(L_{(N,M)}^{k/N})_+,L_{(N,M)}],\\
\pa_k\phi_i &=&((L_{(N,M)}^{k/N})_+\phi_i)_0,
\qquad \pa_k\psi_i=-((L_{(N,M)}^{k/N})^*_+\psi_i)_0
\eea
where $\phi_i$ and $\psi_i$ are eigenfunctions and adjoint eigenfunctions, respectively.
(Notations:$(A)_{\pm}$ denote the differential part and the 
integral part of the pseudo-differential operator 
$A$ respectively, $(A)_0$ denotes the zeroth order term, and * stands for the conjugate
operation: $(AB)^*=B^*A^*,\; \partial^*=-\partial,\; f(x)^*=f(x)$). 

The second Hamiltonian structure associated with $L_{(N,M)}$ 
is given by the second GD bracket as follow
\bea
\Th^{GD}_2(\frac{\de H}{\de L_{(N,M)}})
&=&(L_{(N,M)}\frac{\de H}{\de L_{(N,M)}})_+L_{(N,M)}-
L_{(N,M)}(\frac{\de H}{\de L_{(N,M)}}L_{(N,M)})_+\no\\
& &+\frac{1}{N}[L_{(N,M)},\int^x res[L_{(N,M)},\frac{\de H}{\de L_{(N,M)}}]].
\label{posl1}
\eea
where the last term in (\ref{posl1}) is just the Dirac constraint
 imposed by $u_1=0$ on $L_{(N,M)}$. 

In the following, we will discuss the simplest example
($N=1,M=2$) in detail. 

The 2-constraint KP hierarchy with order one is defined by
\be
L_{(1,2)}=\pa+\phi_1\pari\psi_1+\phi_2\pari\psi_2
\label{laxl1}
\ee
From (\ref{posl1}) the basic second Poisson brackets are given by
\bea
\{\phi_i,\phi_j\}&=&
-(\phi_i\pari\phi_j+\phi_j\pari\phi_i),\no\\
\{\psi_i,\psi_j\}&=&
-(\psi_i\pari\psi_j+\psi_j\pari\psi_i),\\
\{\phi_i,\psi_j\}&=&
(\de_{ij}L_{(1,2)}+\phi_i\pari\psi_j),\no
\eea
which is obviously nonlocal.
The algebraic structure of the Poisson brackets is transparent
if we set $t\equiv \phi_1\psi_1+\phi_2\psi_2$, then
\bea
\{t,t\}&=&2t\pa+t',\no\\
\{\phi_i,t\}&=&\phi_i\pa+\phi_i'
\label{extv1}\\
\{\psi_i,t\}&=&\psi_i\pa+\psi_i'.\no
\eea 
Hence $\phi_i$ and $\psi_i$ are spin-1 fields with respect to the Virasoro
generator $t$, and (\ref{extv1}) form a nonlocal extension of the Virasoro algebra
by four spin-1 fields. We would like to remark that the algebra (\ref{extv1})
can be generalized
to the multi-constraint case ($N=1,M>2$) by setting 
$t=\sum_{i=1}^M\phi_i\psi_i$ .

\section{Nonstandard cKP hierarchy}
The nonstandard Lax operator is obtained by performing 
a gauge transformation on $L_{(1,2)}$ as follow
\bea
K_{(1,2)} &=& \phi^{-1}_1L_{(1,2)}\phi_1,\\
&=& \pa+v_1+\pari v_2+q\pari r
\eea
where
\bea
&& v_1=\phi_1'/\phi_1,\qquad v_2=\phi_1\psi_1,\\
&& q=\phi_1^{-1}\phi_2,\qquad r=\phi_1\psi_2.
\eea
The transformed Lax operator $K_{(1,2)}$ satisfies the hierarchy equations 
\bea
\pa_nK_{(1,2)} &=& [(K_{(1,2)}^n)_{\ge 1},K_{(1,2)}],\no\\
\pa_nq &=&((K_{(1,2)}^n)_{\ge 1}q)_0,
\label{nhe}\\
\pa_nv_2 &=&-((K_{(1,2)}^n)^*_{\ge 1}v_2)_0,\qquad 
\pa_n r=-((K_{(1,2)}^n)^*_{\ge 1}r)_0.\no
\eea
and the transformed  second Hamiltonian structure now becomes\cite{OS}
\bea
\Th^{NS}_2(\frac{\de H}{\de K_{(1,2)}})
&=&(K_{(1,2)}\frac{\de H}{\de K_{(1,2)}})_+K_{(1,2)}-
K_{(1,2)}(\frac{\de H}{\de K_{(1,2)}}K_{(1,2)})_+
+[K_{(1,2)},(K_{(1,2)}\frac{\de H}{\de K_{(1,2)}})]\no\\
& &+\pari res[K_{(1,2)},\frac{\de H}{\de K_{(1,2)}}]K_{(1,2)}
+[K_{(1,2)},\int^x res[K_{(1,2)},\frac{\de H}{\de K_{(1,2)}}]].
\label{nhs}
\eea
where the basic Poisson brackets can be easily written as
\bea
\{v_1,v_1\}&=& 2\pa,\no\\
\{v_1,v_2\}&=&\pa^2+\pa v_1+\pa q\pari r,\no\\
\{v_1,q\}&=&-q'\pari,\no\\
\{v_1,r\}&=&-r,\no\\
\{v_2,v_2\}&=&\pa v_2+v_2\pa+v_2q\pari r+r\pari qv_2,
\label{pov}\\
\{v_2,q\}&=&-\pa q+v_1q-v_2q\pari-r\pari q^2,\no\\
\{v_2,r\}&=&\pa r-v_1r+r\pari qr-r\pari v_2,\no\\
\{q,q\}&=&-2q\pari q+\pari q^2+q^2\pari,\no\\
\{q,r\}&=&\pa+v_1+2q\pari r+\pari v_2-\pari qr,\no\\
\{r,r\}&=&-2r\pari r.\no
\eea
These Poisson brackets are nonlocal as well. To simplified the 
above Poisson brackets, we may consider the operator
\bea
L_{(2,1)} &=&\pa K_{(1,2)}
\label{derk}\\
&=&\pa^2+u_1\pa+u_2+\phi\pa^{-1}\psi
\eea
where
\bea
u_1&=&v_1,\qquad u_2=v_2+v_1'+qr,\no\\
\phi&=&q',\qquad \psi=r.
\label{l2k}
\eea
Using (\ref{pov}) we can calculate the Poisson brackets for
$\{u_1,u_2,\phi,\psi\}$ which now become simpler
\bea
\{u_1, u_1\} &=& 2\pa, \no\\
\{u_1, u_2\} &=& -\pa^2+\pa u_1, \no\\
\{u_1, \phi\} &=& \phi,
\label{brac} \\
\{u_1, \psi\} &=& -\psi, \no\\
\{\phi,\phi\}&=&-2\phi\pa^{-1}\phi\no
\eea
etc. Note that these brackets are not the same as the ones
constructed from the second GD brackets for the 1-constraint
KP hierarchy where the corresponding brackets are given by \cite{OS}
\bea
\{u_1, u_1\} &=& -2\pa, \no\\
\{u_1, u_2\} &=& \pa^2-\pa u_1, \no\\
\{u_1, \phi\} &=& -\phi, \\
\{u_1, \psi\} &=& \psi, \no\\
\{\phi, \phi\} &=& -\phi\pari \phi.\no
\eea
etc. In fact, it can be shown [see Appendix] that (\ref{brac}) 
obey the following Poisson structure
\be
\{F,G\}=
\int res(\frac{\de F}{\de K_{(1,2)}}\Th_2^{NS}(\frac{\de G}{\de K_{(1,2)}}))
=\int res(\frac{\de F}{\de L_{(2,1)}}\Omega(\frac{\de G}{\de L_{(2,1)}}))
\label{id}
\ee
where
\be
\Omega(\frac{\de G}{\de L_{(2,1)}})
=(L_{(2,1)}\frac{\de G}{\de L_{(2,1)}})_+L_{(2,1)}-
L_{(2,1)}(\frac{\de G}{\de L_{(2,1)}}L_{(2,1)})_+
+[L_{(2,1)},\int^x res[L_{(2,1)},\frac{\de G}{\de L_{(2,1)}}]].
\label{pol2}
\ee
Besides the second GD structure, the last term in (\ref{pol2}) 
is called the third GD structure which is
compatible with the second one \cite{DIZ}. Thus the Hamiltonian
structure associated with $L_{(2,1)}$ is the sum of the second and the third
GD structures.

Before ending this section, let us discuss the algebraic structure associated with
the ncKP hierarchy. Based on the dimension consideration, we can define a Virasoro
generator $t\equiv v_2+v_1'/2+qr$. Then from (\ref{pov}), we have
\bea
\{v_1,t\}&=&v_1\pa+v_1' ,\no\\
\{t,t\}&=&\frac{1}{2}\pa^3+2t\pa+t' ,\no\\
\{q,t\}&=&\frac{1}{2}q\pa+q'-\frac{1}{2}\pa^{-1}q\pa^2 ,\no\\
\{r,t\}&=&\frac{3}{2}r\pa+r' .
\label{extv2}
\eea
We see that $v_1$ and $r$ are spin-1 and spin-3/2 fields, respectively and $q$ is not a
spin field due to the anomalus term ``$-\frac{1}{2}\pa^{-1}q\pa^2$ ". However, if we
take a derivative to the third bracket in (\ref{extv2}), then $q'$ becomes a spin-3/2
field, i.e.
\be
\{q',t\}=\frac{3}{2}q'\pa+q''.
\ee
This motivate us to covariantize the Lax operator $L_{(2,1)}$ rather than the
operator $K_{(1,2)}$. Form (\ref{brac}), $L_{(2,1)}$ can be covariantized 
by setting the Virasoro generator $t\equiv u_2-1/2u_1'$, and
\bea
\{u_1,t\}&=&u_1\pa+u_1' ,\no\\
\{t,t\}&=&\frac{1}{2}\pa^3+2t\pa+t' ,\no\\
\{\phi,t\}&=&\frac{3}{2}\phi\pa+\phi' ,\no\\
\{\psi,t\}&=&\frac{3}{2}\psi\pa+\psi' .
\label{extv3}
\eea
Therefore, the conformal algebra associated with $K_{(1,2)}$ is encoded in the
conformal algebra of $L_{(2,1)}$ .

\section{The generalized Miura transformation}

In this section, we will show that the Poisson structure (\ref{pol2}) 
has a very interesting property under factorization of the operator
$L_{(2,1)}$ into multiplication form. Since the operator of the form $L_{(2,1)}$
has multi-boson representations, we can
factorize $L_{(2,1)}$ into the following form
\be
L_{(2,1)}=(\pa-a_1)(\pa-a_2)(\pa-a_3)(\pa-b_1)^{-1}
\label{facl2}
\ee
where the variables $\{u_1,u_2,\phi,\psi\}$ and
$\{a_1,a_2,a_3,b_1\}$ are related by
\bea
u_1&=& b_1-(a_1+a_2+a_3),\no\\
u_2&=&u_1b_1+2b_1'+a_1a_2+a_2a_3+a_1a_3-a_2'-2a_3',\no\\
\phi&=&e^{\int^x b_1}(u_2b_1+u_1b_1'+b_1''-a_1a_2a_3+a_1a_3'+
a_2'a_3+a_2a_3'-a_3''),\no\\
\psi&=&e^{-\int^xb_1}
\eea
which is called the Miura transformation. Now let us first consider
the second GD bracket under the factorization (\ref{facl2}). Thanks to 
the generalized KW theorem \cite{D3,Y,ANP,MR}, 
the second GD bracket can be simplified as
\bea
\{a_i,a_j\}_2^{GD}&=&-\de_{ij}\pa,\no\\
\{b_1,b_1\}_2^{GD}&=&\pa,
\label{2nd}\\
\{a_i,b_1\}_2^{GD}&=&0,\no
\eea 
Hence the remaining tasks are to study the third GD structure. In the
previous paper\cite{ST2}, we have shown that the third structure
 has also a very nice property under factorization of the Lax operator
containing inverse linear terms (\ref{facl2}). It turns out that \cite{ST2}
\be
\{F,G\}^{GD}_3=\int res(\frac{\de F}{\de L_{(2,1)}}[L_{(2,1)},\int^x 
res[L_{(2,1)},\frac{\de G}{\de L_{(2,1)}}]])
=(\sum_{i=1}^3\frac{\de F}{\de a_i}+\frac{\de G}{\de b_1})
(\sum_{j=1}^3\frac{\de F}{\de a_j}+\frac{\de G}{\de b_1})'
\ee
which leads to
\be
\{a_i,a_j\}_3^{GD}=\{a_i,b_1\}_3^{GD}=\{b_1,b_1\}_3^{GD}=\pa.
\label{3rd}
\ee

Combining (\ref{2nd}) with (\ref{3rd}) we obtain
\bea
\{a_i, a_j\} &=& (1-\de_{ij})\pa,\no\\
\{b_1, b_1\} &=& 2\pa, 
\label{pokw}\\
\{a_i, b_1\} &=&\pa. \no
\eea
Therefore, the Lax operator $K_{(1,2)}$ (and hence $L_{(1,2)}$) has a simple 
and local realization of their Poisson structures.

\section{Conclusions}
We have shown that the second Hamiltonian structure of the ncKP
hierarchy has a very simple realization. In terms of
$\{a_1,a_2,a_3,b_1\}$ the Lax operator $K_{(1,2)}$ can be factorized
as
\be
K_{(1,2)}=\pari(\pa-a_1)(\pa-a_2)(\pa-a_3)(\pa-b_1)^{-1}
\ee
and the second Poisson structure (\ref{pov}) is mapped to a much simpler form 
(\ref{pokw}). In general, we should consider the multi-constraint KP 
hierarchy with the Lax operator of the form (\ref{multi}).
After performing the gauge transformation 
$K_{(N,M)}=\phi_1^{-1}L_{(N,M)}\phi_1$, the
Lax operator $L_{(N,M)}$ is transformed to
\be
K_{(N,M)}=\pa^N+v_1\pa^{N-1}+\cdots+v_N+\pa^{-1}v_{N+1}+
\sum_{i=1}^{M-1}q_i\pa^{-1}r_i
\ee
which satisfies the nonstandard hierarchy equations (\ref{nhe}) and 
has the Hamiltonian structure (\ref{nhs}). 
Moreover we can follow the strategy in Appendix to prove without difficulty that 
the Hamiltonian structure associated with the operator 
$L_{(N+1,M-1)}\equiv \pa K_{(N,M)}$ is just the sum of the second 
and third GD structure (\ref{pol2}). Thus by applying the previous results\cite{ST2}, 
the Lax operator of the ncKP hierarchy can be factorized as
\be
K_{(N,M)}=\pa^{-1}(\pa-a_1)\cdots(\pa-a_n)(\pa-b_1)^{-1}\cdots(\pa-b_m)^{-1}
\ee
and the simplified Poisson brackets turn out to be
\bea
\{a_i, a_j\} &=& (1-\de_{ij})\pa,\no\\
\{b_i, b_j\} &=& (1+\de_{ij})\pa, 
\label{bracket}\\
\{a_i, b_j\} &=&\pa. \no
\eea
Finally we would like to remark that the Poisson bracket matrix (\ref{bracket})
is symmetric and nonsingular, thus it is not difficult to diagonalize the
matrix to obtain the free field representation which would be useful to
quantize the W-algebra associated with the ncKP hierarchy. 
The details of these discussions will be presented 
in a forthcoming paper\cite{HST}.

{\bf Acknowledgments\/}
We would like to thank Professors J-C Shaw and W-J Huang for inspiring discussions
and Dr. M-C Chang for reading the manuscript. 
This work is supported by the National Science Council of the
Republic of China under grant No. NSC-86-2112-M-007-020.

\appendix
\section{}
In this appendix we give a proof of (\ref{id}).
From (\ref{l2k}) we have
\bea
\frac{\de H}{\de v_1}&=&\frac{\de H}{\de u_1}-(\frac{\de H}{\de u_2})',
\qquad \frac{\de H}{\de v_2}=\frac{\de H}{\de u_2},
\label{ch1}\\
\frac{\de H}{\de q}&=&r\frac{\de H}{\de v_2}-(\frac{\de H}{\de \phi})',
\qquad \frac{\de H}{\de r}=q\frac{\de H}{\de v_2}+
\frac{\de H}{\de \psi}.
\label{ch2}
\eea
Let 
\be
\frac{\de H}{\de K_{(1,2)}}=\pari \frac{\de H}{\de v_1}+
\frac{\de H}{\de v_2}+A
\ee
where $A=(A)_{\ge 0}$. Then from
\be
\de H=\int res(\frac{\de H}{\de K_{(1,2)}}\de K_{(1,2)})
=\int(\frac{\de H}{\de v_1}\de v_1+\frac{\de H}{\de v_2}\de v_2
+\frac{\de H}{\de q}\de q+\frac{\de H}{\de r}\de r)
\ee
we have
\bea
(A)_0 &=&0\\
(Aq)_0 &=&\frac{\de H}{\de r}-q\frac{\de H}{\de v_2}=\frac{\de H}{\de \psi},\\
(A^*r)_0 &=&\frac{\de H}{\de q}-r\frac{\de H}{\de v_2}=
-(\frac{\de H}{\de \phi})'.
\eea
Note that $A$, in fact, is a pure differential operator.
Now from (\ref{ch1}) and (\ref{ch2}) 
\bea
\frac{\de H}{\de K_{(1,2)}}\pari
&=&(\pari \frac{\de H}{\de v_1}+\frac{\de H}{\de v_2}+A)\pari \no\\
&=&(\pari\frac{\de H}{\de v_2}+\pa^{-2}(\frac{\de H}{\de v_1}
+(\frac{\de H}{\de v_2})')+A\pari)+O(\pa^{-3})\no\\
&=&(\pari\frac{\de H}{\de u_2}+\pa^{-2}\frac{\de H}{\de u_1}
+A\pari)+O(\pa^{-3})\no\\
&=&(\frac{\de H}{\de L_{(2,1)}})_-+A\pari+O(\pa^{-3}).
\label{}
\eea
Let us define $B=A\pari$, then
\be
(B\phi)_0=(A\pari \phi)_0
=(Aq)_0
=\frac{\de H}{\de \psi}.
\label{b1}
\ee
On the other hand,
\be
(B^*\psi)_0=-(\pari A^*r)_0
=-\int^x(A^*r)_0
=\frac{\de H}{\de \phi}.
\label{b2}
\ee
Eqs.(\ref{b1}) and (\ref{b2}) imply that
\be
B=A\pari=(\frac{\de H}{\de L_{(2,1)}})_+
\ee
and hence
\be
\frac{\de H}{\de K_{(1,2)}}\pari=\frac{\de H}{\de L_{(2,1)}}+O(\pa^{-3}).
\label{id2}
\ee
Combining (\ref{derk}) and (\ref{id2}),  it is easy to derive \cite{HSY} 
the relation (\ref{id}).

\end{document}